\documentclass[
  reprint,
  superscriptaddress,
  amsmath,amssymb,
  aps,
  prl,
  showpacs,
  floatfix
]{revtex4-2}
\usepackage{graphicx}   
\usepackage{dcolumn}    
\usepackage{bm}         
\usepackage{hyperref}   
\usepackage[percent]{overpic}

\begin{document}

\title{Light induced superconducting diode effect in patterned films}

\author{Evan M. Wilson}
\affiliation{Department of Physics, University of Connecticut, Storrs, Connecticut 06269, USA}

\author{Hou-Tong Chen}
\affiliation{Center for Integrated Nanotechnologies, Los Alamos National Laboratory, Los Alamos, New Mexico, USA}

\author{Alexander V. Balatsky}
\affiliation{Department of Physics, University of Connecticut, Storrs, Connecticut 06269, USA}
\affiliation{Nordita, KTH Royal Institute of Technology and Stockholm University, SE-106 91 Stockholm, Sweden}

\date{\today}

\begin{abstract}
Structured light offers a route to control superconducting transport without
permanently modifying the material or applying a static bias. Here we show
that structured optical driving can generate a superconducting diode response
in patterned superconducting films with asymmetric holes. Using generalized
time-dependent Ginzburg--Landau simulations, we find that optical driving
produces rectified dc photovoltages and zero-bias directional supercurrent
imbalance in a junction-free geometry, with continuous-drive diode efficiencies
of order $10^{-3}$ and pulsed efficiencies reaching $10^{-2}$. The response is
controlled by both the hole array and the optical mode. Increasing the number
of asymmetric holes enhances rectification, reversing circular helicity reverses
the diode polarity, and the optical spatial mode strongly modifies the magnitude
and polarity of the directional response. Pulsed excitation enhances the
zero-bias line-cut current imbalance to the percent level. For linearly polarized
illumination, the asymmetric metacrystal converts the drive into local chiral
supercurrent motion, inducing an inverse-Faraday-effect-like mechanism for
dynamical time-reversal-symmetry breaking. These results establish patterned
superconducting films as a viable platform for light-tunable superconducting
diode behavior.
\end{abstract}

\maketitle

\section{Introduction} 

Since the first experimental realization of the superconducting diode effect (SDE) in the noncentrosymmetric superlattice $[{\rm Nb/V/Ta}]_n$ \cite{Ando2020Nature}, nonreciprocal superconducting transport has become a rapidly growing area of research. Subsequent work showed that diode behavior can arise through several avenues such as intrinsic noncentrosymmetric superconductivity, finite-momentum pairing, Josephson structures, vortex-limited critical currents, magnetochiral anisotropy, and geometrically patterned films \cite{Daido2022PRL,Baumgartner2022JosephsonDiode,Yuan2022PNAS,Lyu2021NatCommun,Suri2022APL,Pal2022NatPhys,Chen2018PRB,Legg2022PRB}. In gate-tunable InAs/Al interferometers, a tunable Josephson diode effect arises from finite flux bias, nonsinusoidal current-phase relations, and SQUID-arm asymmetry \cite{Ciaccia2023GateTunableJD}. Conventional tunnel-junction platforms can also depend sensitively on the microscopic barrier structure, including oxygen deficiency and interface disorder in Al/AlO$_x$/Al junctions \cite{Zeng2016AlOxBarrier}. More recently, twisted high-$T_c$ cuprate Josephson devices have been shown to achieve perfect diode efficiency under microwave irradiation above liquid-nitrogen temperature \cite{Wang2025PerfectQSD}.

Early theories of polar superconductors established that broken
inversion symmetry permits magnetoelectric coupling and additional
Ginzburg--Landau terms that can produce direction-dependent critical
currents in a magnetic field
\cite{Edelstein1995PRL,Edelstein1996JPCM}. The intrinsic SDE requires inversion symmetry breaking together with time-reversal symmetry breaking, so that opposite transport directions are no longer symmetry equivalent\cite{Nadeem2023NatRevPhys}. This setting also permits second-order optical response, which makes optical rectification a new route to superconducting diode physics. It has been found that irradiation can generate a phase accumulation of dc superconducting condensate in superconductors with diode electrodynamics. Related proposals show that electromagnetic driving can produce a photodiode-like superconducting response controlled by the polarization and frequency of the incident field \cite{Wakatsuki2017SciAdv,Mironov2024PRB,Parafilo2025Photodiode,Nagaosa2024ARCMP}. 

Directional photocurrents have also been observed in metamaterials and symmetry-broken metasurfaces that support light-driven vectorial currents \cite{Matsubara2022Polarization,Pettine2024LightDriven,Wei2020ZeroBiasMIR,Wei2023GeometricMIR}. Related work indicates that linearly polarized light can drive local elliptical charge motion with a finite transverse phase lag in structurally asymmetric nano-antennas~\cite{Yang2023LinearIFE}. Linearly polarized induced orbital magnetization mechanisms, including orbital inverse Faraday and inverse Cotton--Mouton responses, have been discussed in driven superconductors and related charged fluid~\cite{Mironov2021IFECondensates, Cordoso2026PRL}. 

Linearly polarized optical drives have been demonstrated to imprint nontrivial superconducting vorticity through its spatially structured vector potential \cite{Yeh2025QuantumPrintingI}. This idea was subsequently extended to Laguerre-Gaussian beams, where the spin angular momentum (SAM), orbital angular momentum (OAM), and radial structure of the optical mode provide additional control over the dynamics of the induced vorticity \cite{Yeh2025QuantumPrintingII}. 

\begin{figure}[t]
    \centering
    \includegraphics[width=0.95\columnwidth]{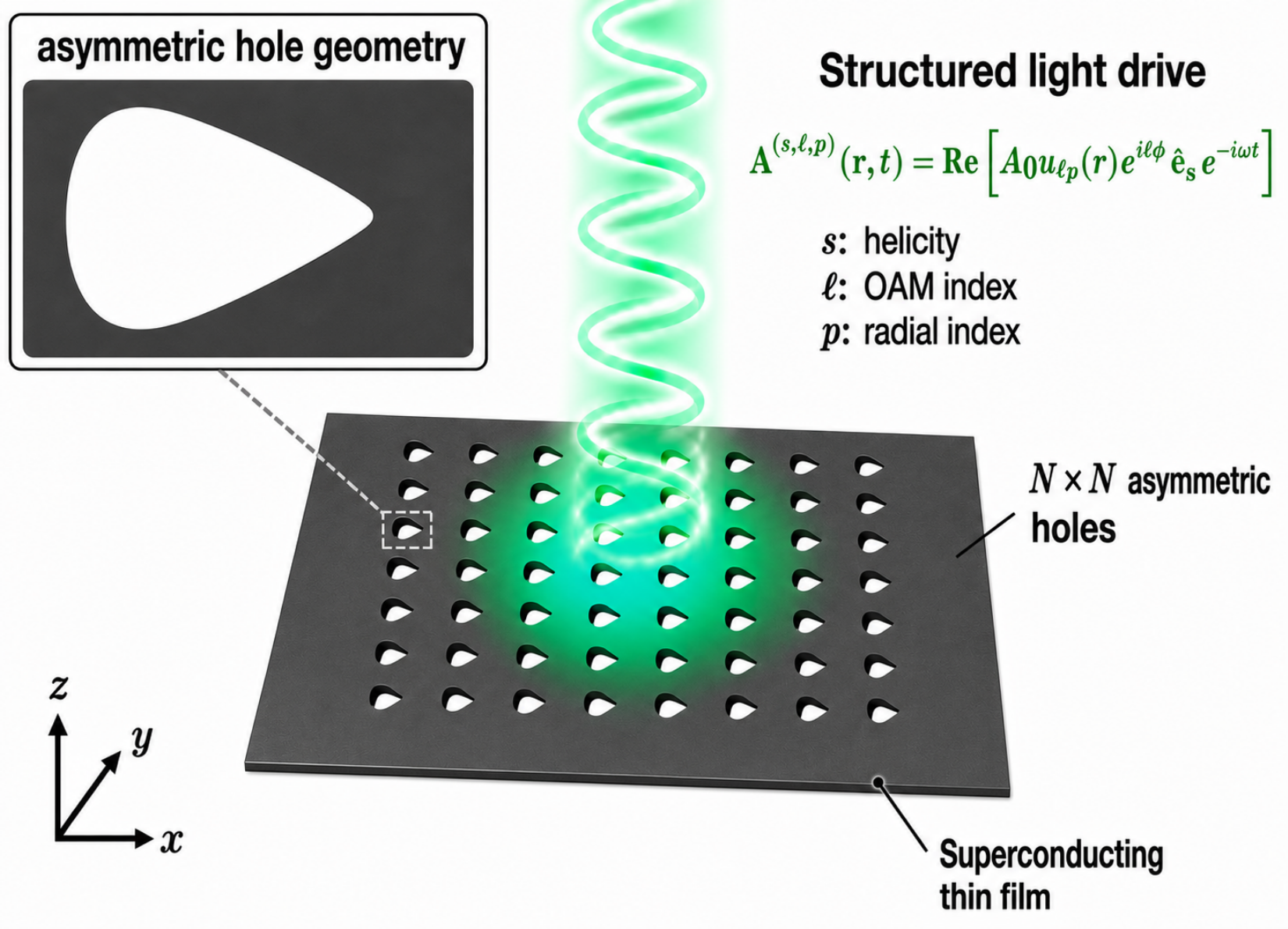}
    \caption{\raggedright 
    A superconducting film patterned with an $N_x \times N_y$ array of asymmetric holes is illuminated by a structured optical beam. 
    The drive is described by a vector potential $\mathbf{A}^{(s,\ell,p)}(\mathbf{r},t)$, where $s$ denotes helicity, $\ell$ the orbital angular momentum index, and $p$ the radial index. The induced condensate dynamics are modeled using the time-dependent Ginzburg--Landau equation, with the supercurrent density $\mathbf{j}_s$ determined by the gauge-covariant phase gradients of the superconducting order parameter.
    }
    \label{fig:structured_light_tdgl_schematic}
\end{figure}

In this work, we adopt the quantum-printing perspective, where structured photons provide a route to imprint nonequilibrium states and topological excitations onto quantum matter \cite{Aeppli2025QuantumPrinting}. Using generalized time-dependent Ginzburg--Landau theory \cite{KramerWattsTobin1978,BISHOP23CPC}, we study patterned superconducting thin films perforated by geometrically asymmetric holes and show that THz-frequency structured light generates finite cycle-averaged currents and photovoltages in a junction-free geometry. The optical mode parameters, including the helicity/SAM index $s$, OAM index $\ell$, and radial order $p$, act as control parameters for the rectified condensate flow written into the superconducting film. The patterned-film geometry and structured optical drive are illustrated in Fig.~\ref{fig:structured_light_tdgl_schematic}.

We find that the response contains a leading rectification component together with higher-order nonlinear corrections at stronger drive, and is affected by the number of asymmetric holes and their lattice geometry. The dc response is further controlled by the optical mode structure, with helicity producing a sign-selective response and OAM modifying the
magnitude and direction of the asymmetry. Our results establish patterned superconducting films as a platform for optically controlled nonreciprocal transport and suggest the path toward a light-tunable superconducting diode effect. A realistic THz simulation involves a large separation
between the electromagnetic wavelength, which may be hundreds of
micrometers to millimeters, and the superconducting coherence length.
The corresponding antenna and patterned metacrystal would therefore
extend over similarly large dimensions. Resolving these electromagnetic
scales together with coherence length scale condensate dynamics is
computationally demanding, so we use reduced optical and sample
dimensions to isolate the rectification mechanism. These calculations
provide a proof of principle; future work will combine full-wave
simulations of a realistic THz antenna and its near field with TDGL
simulations of the superconducting response.

\section{Simulation Methods}

We model the driven superconducting film using the generalized time-dependent Ginzburg--Landau equation for a thin superconducting film,
\begin{equation}
\begin{aligned}
&\frac{u}{\sqrt{1+\gamma^2|\psi|^2}}
\left(
\partial_t+i\mu+\frac{\gamma^2}{2}\partial_t|\psi|^2
\right)\psi
\\
&=
(1-|\psi|^2)\psi
+
(\nabla-i\mathbf A)^2\psi .
\end{aligned}
\label{eq:gtdgl}
\end{equation}
Here \(\psi(\mathbf r,t)=|\psi(\mathbf r,t)|e^{i\theta_s(\mathbf r,t)}\) is the complex superconducting order parameter, with phase field \(\theta_s\). The electrochemical potential is denoted by
\(\mu(\mathbf r,t)\), and \(\mathbf A(\mathbf r,t)\) is the vector potential. The parameter \(u=5.79\) sets the relaxation time scale of the order parameter relative to the supercurrent response, while \(\gamma=10\) accounts for inelastic scattering effects. The total current density is written as the sum of superconducting and normal contributions, $J=J_s+J_n$ with
\begin{equation}
\mathbf J_s
=
{\rm Im}
\left[
\psi^*(\nabla-i\mathbf A)\psi
\right],
\qquad
\mathbf J_n
=
-\sigma
\left(
\nabla\mu+\partial_t\mathbf A
\right).
\label{eq:currents}
\end{equation}
For the thin-film geometry, we denote the corresponding sheet-current density by
\(\mathbf{K}=d\mathbf{J}\), where \(d\) is the film thickness. In dimensionless units with \(\sigma=1\), this reduces to \(\mathbf J_n=-\nabla\mu-\partial_t\mathbf A\).

We define the system as a square superconducting thin film perforated by geometrically asymmetric holes. We consider both a single-hole device and \(N_x\times N_y\) arrays of identical asymmetric holes with different array sizes and spatial arrangements. The holes have a smooth, guitar-pick-like shape that breaks inversion symmetry, while the superconducting material itself is taken to be homogeneous. The electrochemical potential is determined self-consistently by charge conservation with insulating gauge-covariant boundary conditions on the outer film edges and hole boundaries. Magnetic self-field screening is neglected, so the vector potential is taken to be the externally applied optical vector potential. The explicit boundary conditions, material and lattice parameters, hole-number scan geometry, and convergence tests are given in Appendix~\ref{app:numerics}.

\begin{figure}[t]
    \centering

    \includegraphics[width=\columnwidth]{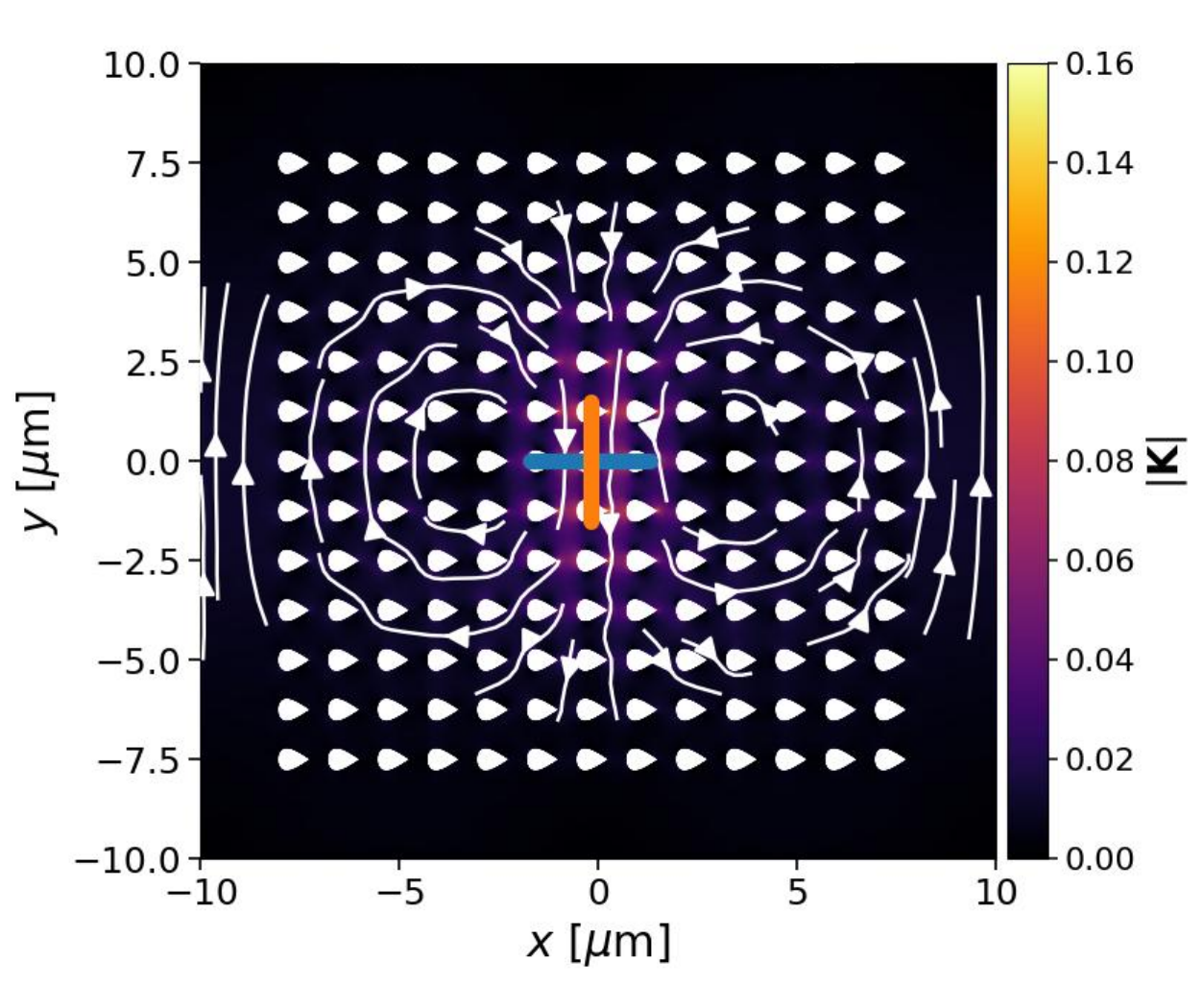}

    \caption{\raggedright
    Optically induced sheet-current texture in the 169-hole superconducting metacrystal for the continuous linearly polarized Gaussian drive, \((s,\ell,p)=(0,0,0)\). The color scale shows the magnitude of the induced current, and the horizontal and vertical central cuts mark the internal channels used to compare left-right and up-down current flow. The asymmetric hole array redirects the driven current into components parallel and perpendicular to the driving electric field, enabling a zero-bias directional current imbalance. Video simulations for all optical modes are available on the companion website~\cite{MetaSDEVideos2026}.
    }
    \label{fig:supercurrent_density}
\end{figure}

The optical drive is introduced through a time-dependent vector potential
\(\mathbf A_{\rm opt}^{(s,\ell,p)}(\mathbf r,t)\), which enters the TDGL
equation through the gauge-covariant derivative \(\nabla-i\mathbf A\). We
write the drive in the general form
\begin{equation}
\mathbf A_{\rm opt}^{(s,\ell,p)}(\mathbf r,t)
=
{\rm Re}
\left[
A_0 u_{\ell p}(r)
e^{i\ell\varphi}
\hat{\mathbf e}_s
e^{-i\omega t+i\phi_0}
\right],
\label{eq:optical_drive}
\end{equation}
where \(A_0=E_0/\omega\), \(u_{\ell p}(r)\) is the transverse spatial mode profile, \(\ell\) is the orbital angular momentum index, \(p\) is the radial index, and \(\hat{\mathbf e}_s\) defines the polarization state. Unless otherwise stated, we use \(E_0=0.5\), beam waist \(w_0=3~\mu{\rm m}\), phase \(\phi_0=0\), and dimensionless drive frequency \(\omega=2\pi/10\), corresponding to a period \(T=10\). The modes \((0,0,0)\) and \((0,1,0)\) denote linearly \(y\)-polarized Gaussian and Laguerre--Gaussian beams, respectively, while \((\pm1,0,0)\) and \((\pm1,1,0)\) denote circular Gaussian and circular Laguerre--Gaussian beams with opposite helicities. Physical-unit conversion, pulse-envelope parameters, polarization normalization, and the \(x\)-polarized control calculation are given in Appendix~\ref{app:numerics}.

The dc photovoltage is obtained from the cycle-averaged electrochemical-potential difference between left and right probes at \((x,y)=(-9~\mu{\rm m},0)\) and \((9~\mu{\rm m},0)\), respectively, outside the patterned region as discussed in Appendix~\ref{app:numerics}.

To characterize the internal directional current imbalance, we integrate the sheet-current density through line cuts placed at the center of the film. For a cross section \(C\), the current is determined as
\(I_C=\int_C \mathbf{K}(\mathbf r,t)\cdot\hat{\mathbf n}\,dl\), where
\(\hat{\mathbf n}\) is the normal direction of the chosen cut. Horizontal and vertical cuts passing through the center of the film define the current
channels up, down, left, and right. We define the corresponding current imbalance as

\begin{equation}
\eta_{\rm LR}
=
\frac{I_{\rm left}-I_{\rm right}}
{I_{\rm left}+I_{\rm right}},
\qquad
\eta_{\rm UD}
=
\frac{I_{\rm up}-I_{\rm down}}
{I_{\rm up}+I_{\rm down}} .
\label{eq:eta}
\end{equation}

These coefficients quantify the directional asymmetries of the internal coherent sheet-current flow through the central line cuts.

In the illuminated patterned films, an additional dc bias nucleates vortices
and phase slips, so critical-current sweeps would be dominated by dissipative
vortex motion. The line cut definition isolates the directional imbalance of the coherent sheet current before bias-driven dissipation sets in.

\section{Directional Supercurrent Asymmetry}

To understand how a linearly polarized drive can generate a diode-like
response, we examine the symmetry of the Gaussian mode as it interacts with the inversion-broken patterned film. For the Gaussian mode
\((s,\ell,p)=(0,0,0)\), we choose a standard \(y\)-polarized beam such that
\begin{figure}[t]
    \centering

    \begin{overpic}[
        width=\columnwidth,
        height=0.68\columnwidth,
        keepaspectratio=false
    ]{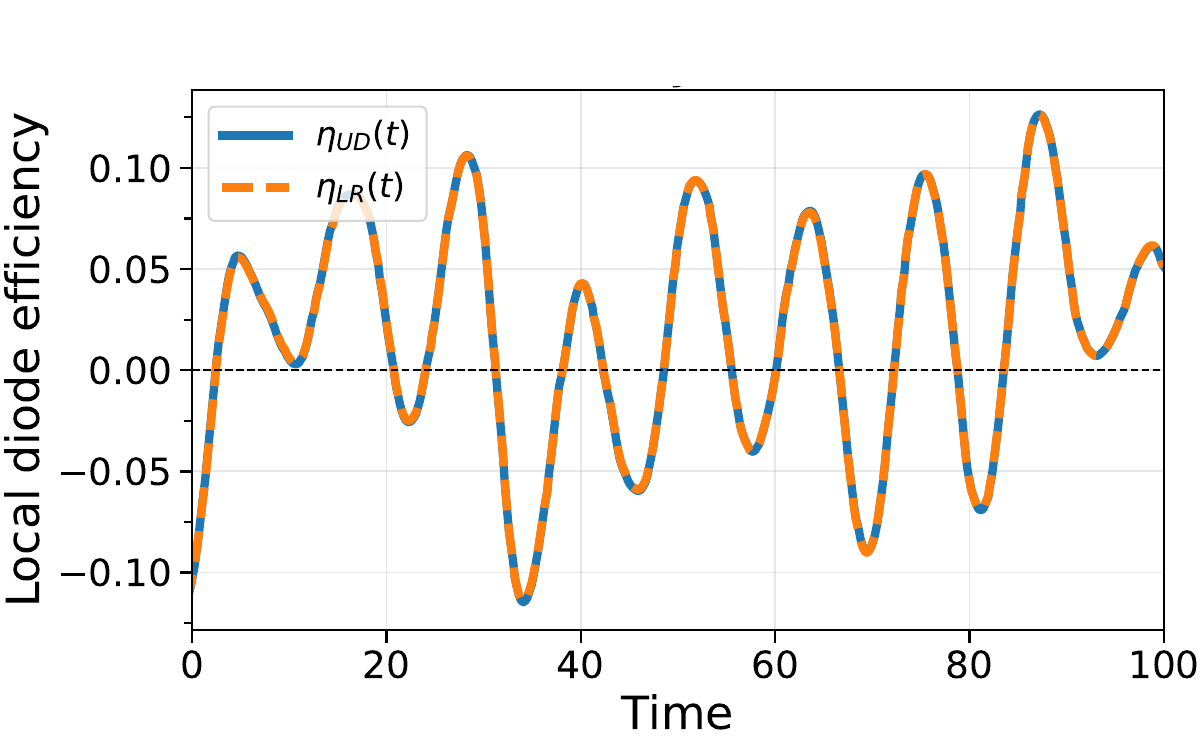}
        \put(1,60){(a)}
    \end{overpic}

    \vspace{0.15cm}

    \begin{overpic}[
        width=\columnwidth,
        height=0.68\columnwidth,
        keepaspectratio=false
    ]{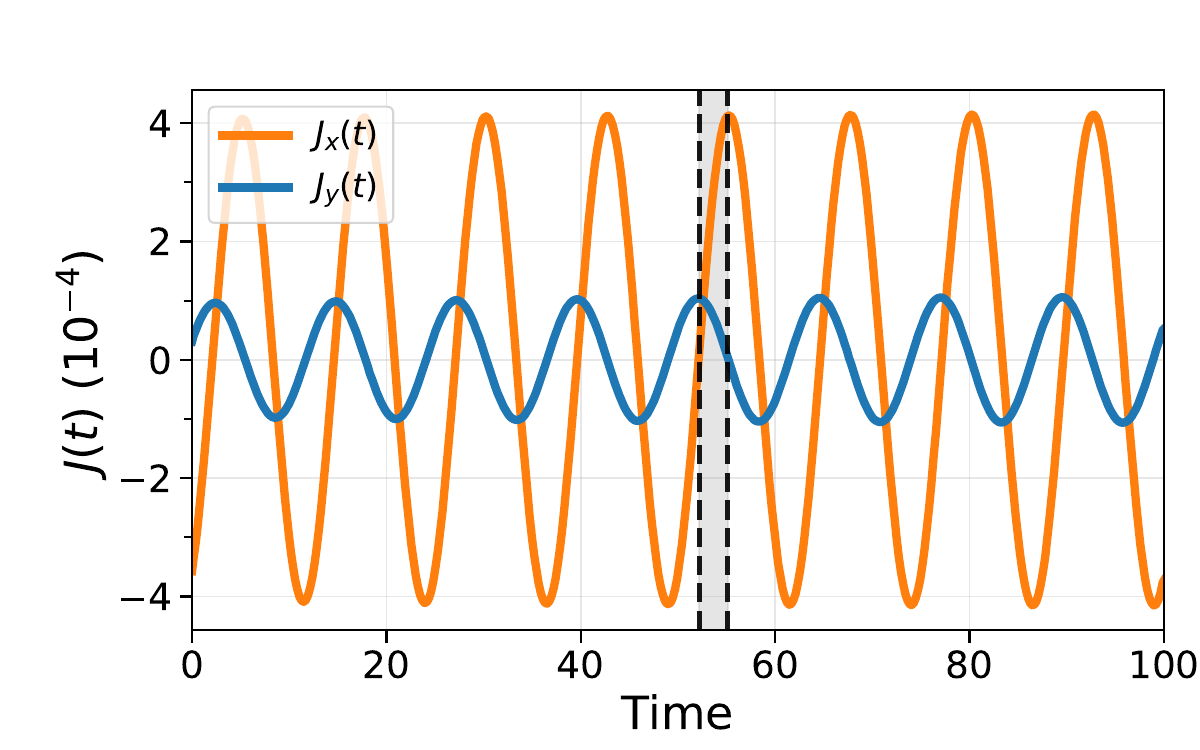}
        \put(1,60){(b)}
    \end{overpic}
    \caption{\raggedright
Time-dependent zero-bias current imbalance and local phase mixing in the 169-hole superconducting metacrystal under continuous linearly polarized
Gaussian illumination, \((s,\ell,p)=(0,0,0)\). Panel (a) shows the
instantaneous left--right and up--down diode-like coefficients,
\(\eta_{\rm LR}\) and \(\eta_{\rm UD}\), obtained from opposite central
line-cut currents. The instantaneous coefficients are plotted as dimensionless
ratios, so \(\eta=0.10\) corresponds to \(10\%\); Table~\ref{tab:eta_modes}
reports the cycle-averaged values in percent. The asymmetric peaks over an optical cycle produce a finite
cycle-averaged directional imbalance. Panel (b) shows representative local
current components \(J_x(t)\) and \(J_y(t)\) inside the patterned region. The
finite phase lag between the two components indicates local elliptical current
motion generated by the asymmetric-hole metacrystal.
    }
    \label{fig:eta_phase_lag}
\end{figure}

\begin{equation}
A_y(\mathbf r,t)=A_0 g_{00}(\mathbf r)\cos(\omega_0 t),
\qquad A_x=0 .
\end{equation}

This drive does not explicitly break time-reversal symmetry. The
vector potential satisfies

\begin{equation}
-\mathbf A(\mathbf r,-t+T/2)=\mathbf A(\mathbf r,t),
\qquad T=\frac{2\pi}{\omega_0},
\end{equation}
and is invariant under the combined operation of time reversal $\mathcal T$ and shift of time variable by $T/2$ labeled $\tau_{T/2}$: 
\begin{equation}
\tilde{\mathcal T}=\tau_{T/2}\mathcal T .
\end{equation}

The diode-like response therefore comes from the driven superconducting state. Dynamical time-reversal-symmetry breaking within the superconducting state is identified by the
failure of the current to map onto its half-period time-reversed
partner such that
\begin{equation}
\mathbf{J}(\mathbf{r},t)
\neq
-\mathbf{J}(\mathbf{r},-t+T/2).
\end{equation}

To characterize the accompanying local current phase mixing, we
analyze the first-harmonic current response. Writing the current
component at $\omega_0$ as $J_i^{(1)}(\mathbf{r})$, we define
\begin{equation}
\chi_j(\mathbf{r})
=
\operatorname{Im}
\left[
J_x^{(1)}(\mathbf{r})
J_y^{(1)*}(\mathbf{r})
\right].
\end{equation}
A finite $\chi_j(\mathbf{r})$ indicates a relative phase between
$J_x^{(1)}$ and $J_y^{(1)}$, so that the local current trajectory
becomes elliptical.

The current response is shown in Figs.~\ref{fig:supercurrent_density} and~\ref{fig:eta_phase_lag}. Figure~\ref{fig:supercurrent_density} displays the optically induced supercurrent density texture. The internal line cuts provide a way to quantify the redistribution of current inside the illuminated metacrystal. The applied vector potential is polarized along a single direction, but the asymmetric hole array converts the drive into a spatially structured current pattern with components parallel and perpendicular to the driving electric field.

The zero-bias diode coefficients for the linear polarized case are shown in Fig.~\ref{fig:eta_phase_lag}(a) and summarized along with other optical modes in Table~\ref{tab:eta_modes}. \(\eta_{\rm LR}\) and \(\eta_{\rm UD}\) show comparable directional redistribution through the two central line cuts for this arrangement. Across the other optical modes, the continuous-drive coefficients remain sub-percent, reversing circular helicity reverses the diode polarity, and pulsed excitation enhances several coefficients to the percent level. The linearly polarized Laguerre--Gaussian pulse gives the largest up--down response, \(\eta_{\rm UD}=1.28\%\). Detailed mode-by-mode trends are given in Appendix~\ref{app:modes}.

A microscopic origin of this directional asymmetry is shown in
Fig.~\ref{fig:eta_phase_lag}(b), where the local current components
\(J_x(t)\) and \(J_y(t)\) inside the metacrystal acquire a finite phase lag. The phase lag is evaluated inside the patterned region because this is where the asymmetric holes locally mix the driven current components giving a non-zero \(\chi\). By contrast, probing points outside the patterned region produces no comparable phase lag, indicating that the effect is generated by the metacrystal. The local
current response becomes elliptically polarized within the metacrystal. The same mechanism appears for the linearly polarized Laguerre--Gaussian drive, whereas the corresponding phase-mixed response is strongly suppressed for circular polarization. We interpret this behavior as evidence for an inverse-Faraday-effect-like mechanism in which the asymmetric holes act as complementary nanoantennas, as suggested by Babinet’s principle, converting the linearly polarized drive into local chiral current motion analogous to the elliptical near field motion and inverse Faraday response generated in asymmetric metallic nanoantennas~\cite{VanDerZiel1965,Yang2023LinearIFE}.

\begin{table*}[t]
\caption{
Zero-bias optical diode coefficients for continuous and pulsed drives in
patterned superconducting films. The diode efficiency coefficients \(\eta_{\rm LR}\) and
\(\eta_{\rm UD}\) quantify the normalized imbalance between integrated sheet
currents through opposite central line cuts, with \(\eta_{\rm LR}=(I_{\rm left}-I_{\rm right})/(I_{\rm left}+I_{\rm right})\)
and \(\eta_{\rm UD}=(I_{\rm up}-I_{\rm down})/(I_{\rm up}+I_{\rm down})\).
Positive and negative values indicate opposite optical-diode polarities.
Continuous-drive data are shown for 1, 56, and 169 asymmetric holes, while
pulsed-drive data are shown for the 169-hole metacrystal.
}
\label{tab:eta_modes}
\begin{ruledtabular}
\begin{tabular}{lcccccccc}
\((s,\ell,p)\)
& \multicolumn{2}{c}{1 hole}
& \multicolumn{2}{c}{56 holes}
& \multicolumn{2}{c}{169 holes}
& \multicolumn{2}{c}{169 holes, pulsed} \\
\cline{2-3}
\cline{4-5}
\cline{6-7}
\cline{8-9}
& \(\eta_{\rm LR}\) & \(\eta_{\rm UD}\)
& \(\eta_{\rm LR}\) & \(\eta_{\rm UD}\)
& \(\eta_{\rm LR}\) & \(\eta_{\rm UD}\)
& \(\eta_{\rm LR}\) & \(\eta_{\rm UD}\) \\
\hline
\((0,0,0)\)
& \(-0.12\%\) & \(-0.13\%\)
& \(-0.17\%\) & \(-0.17\%\)
& \(-0.24\%\) & \(-0.24\%\)
& \(-1.145\%\) & \(-1.13\%\) \\

\((+1,0,0)\)
& \(0.08\%\) & \(0.08\%\)
& \(0.13\%\) & \(0.14\%\)
& \(0.26\%\) & \(0.27\%\)
& \(1.21\%\) & \(1.00\%\) \\

\((-1,0,0)\)
& \(-0.08\%\) & \(-0.08\%\)
& \(-0.14\%\) & \(-0.14\%\)
& \(-0.26\%\) & \(-0.27\%\)
& \(-1.19\%\) & \(-1.00\%\) \\

\((0,1,0)\)
& \(0.14\%\) & \(0.17\%\)
& \(0.23\%\) & \(0.33\%\)
& \(0.24\%\) & \(0.54\%\)
& \(0.91\%\) & \(1.28\%\) \\

\((+1,1,0)\)
& \(-0.07\%\) & \(-0.08\%\)
& \(-0.14\%\) & \(-0.17\%\)
& \(-0.14\%\) & \(-0.17\%\)
& \(-0.95\%\) & \(-0.94\%\) \\

\((-1,1,0)\)
& \(0.07\%\) & \(0.08\%\)
& \(0.08\%\) & \(0.09\%\)
& \(0.33\%\) & \(0.39\%\)
& \(1.13\%\) & \(0.88\%\) \\
\end{tabular}
\end{ruledtabular}
\end{table*}

Extended current textures comparing the single-hole and 169-hole geometries across the optical modes are shown in Fig.~\ref{fig:current_textures} and discussed in Appendix~\ref{app:modes}.

\section{DC rectification}

\begin{figure}[t]
    \centering

    \begin{overpic}[width=0.96\columnwidth]{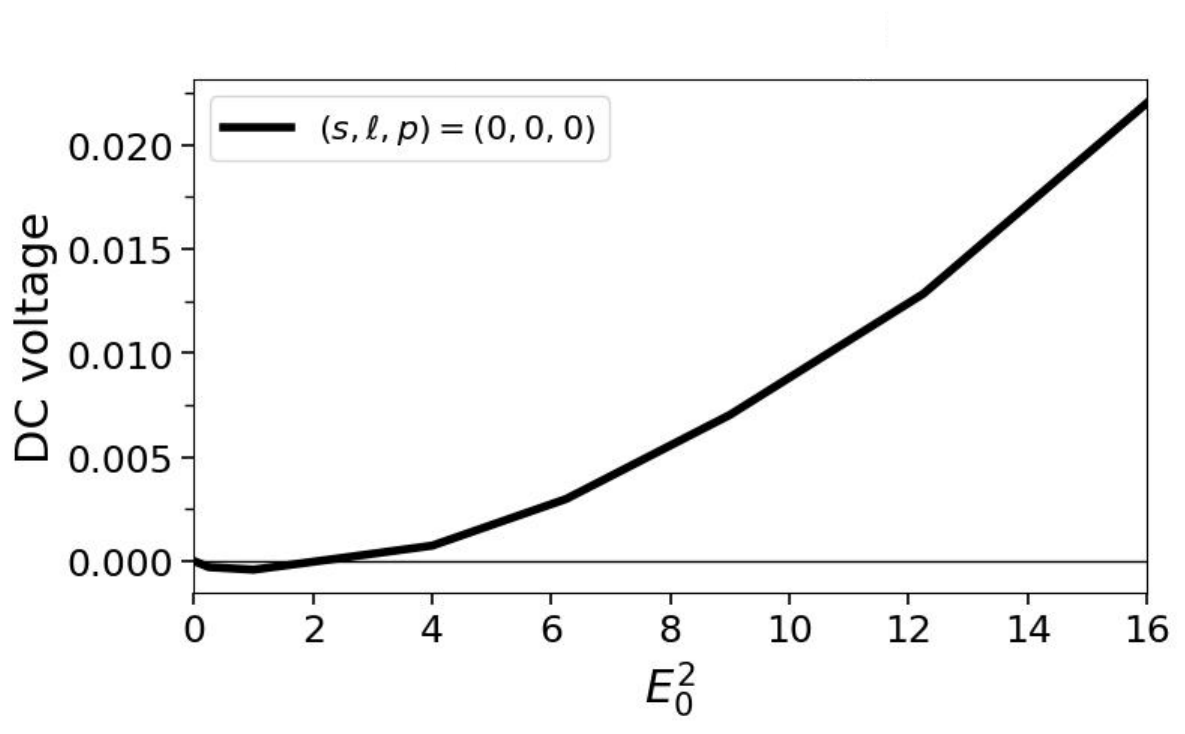}
        \put(-2,55){(a)}
    \end{overpic}

    \vspace{0.35em}

    \begin{overpic}[width=0.96\columnwidth]{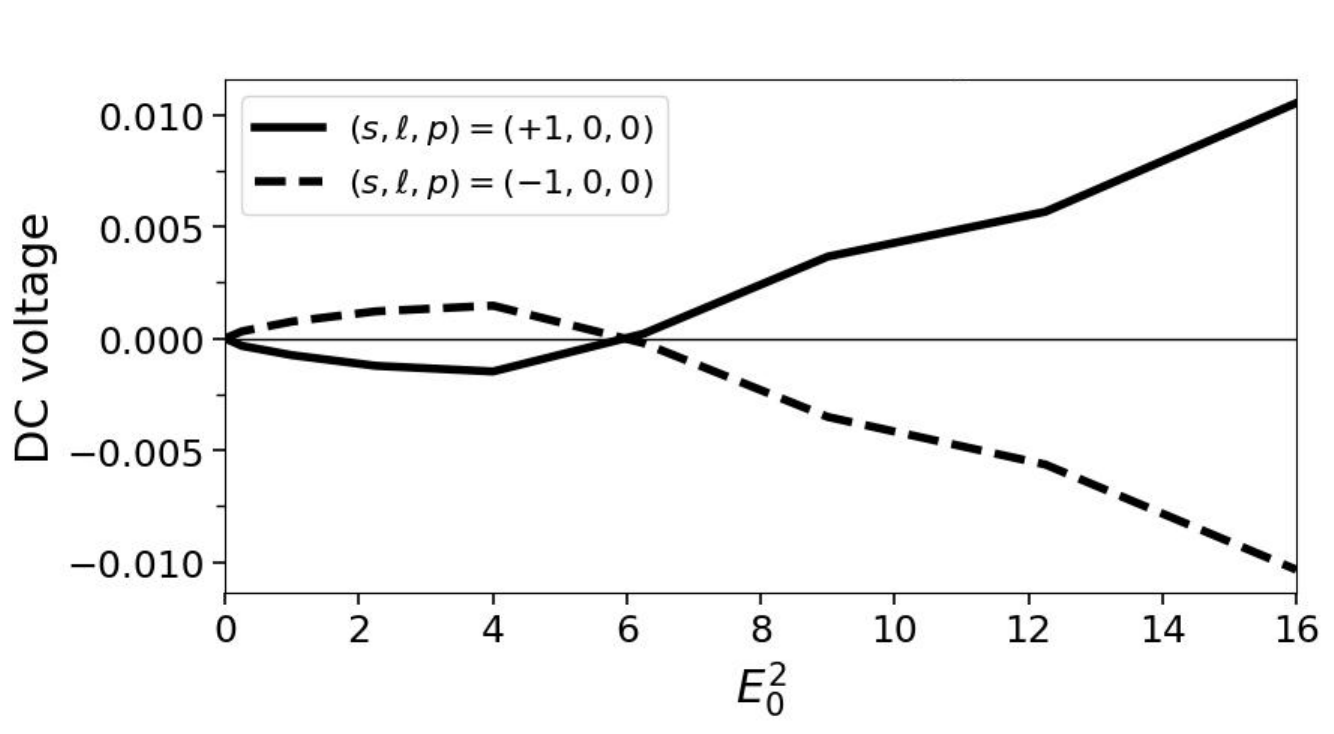}
        \put(-2,55){(b)}
    \end{overpic}

    \caption{\raggedright
    (a) For Gaussian linear polarization, \((s,\ell,p)=(0,0,0)\), the dc voltage
shows nonlinear rectification as a function of the optical intensity proxy
\(E_0^2\). The response includes a sign change at low intensity and higher-order nonlinearity.
(b) Circularly polarized beams with opposite optical helicities, \((s,\ell,p)=(+1,0,0)\) and \((-1,0,0)\), separate into opposite-sign voltage branches that both change sign, showing helicity-sensitive rectification.
    }
    \label{fig:dc_voltage_summary}
\end{figure}

We focus on the Gaussian \(y\)-polarized drive,
\((s,\ell,p)=(0,0,0)\), as the minimal case for demonstrating optical
rectification. The asymmetric-hole geometry breaks left--right reflection
symmetry under \(x\rightarrow -x\), allowing a macroscopic rectified
photovoltage along the \(x\) direction. The \(y\)-polarized drive produces
both \(J_x\) and \(J_y\) through geometry-induced current mixing, and the dc
photovoltage is evaluated between the left and right probes.

Since this drive carries no helicity or orbital angular momentum, the
resulting dc voltage directly isolates the role of the nonlinear
superconducting response. Since the optical intensity is proportional to the square of the electric-field amplitude, Fig.~\ref{fig:dc_voltage_summary}(a) plots the linearly polarized Gaussian response versus \(E_0^2\). The voltage shows a weak negative response
at low intensity, followed by a sign change and rapid positive growth at
larger drive. This behavior is described by higher-order nonlinear terms of
the form \(V_{\rm dc}=aE_0^2+bE_0^4+cE_0^6+\cdots\).

For circular polarization, Fig.~\ref{fig:dc_voltage_summary}(b), the response is strongly helicity dependent. Opposite helicities generate voltage branches with opposite signs at larger intensity, with each branch also showing a low-intensity sign reversal. The curvature and sign changes in both panels indicate that the optical drive pushes the superconducting film beyond the weak drive rectification regime. The nonlinear dc voltage is attributed to higher-order supercurrent response, order-parameter deformation, and geometry-induced current redistribution around the asymmetric holes.

\begin{figure}[t]
    \centering

    \begin{overpic}[width=0.96\columnwidth]{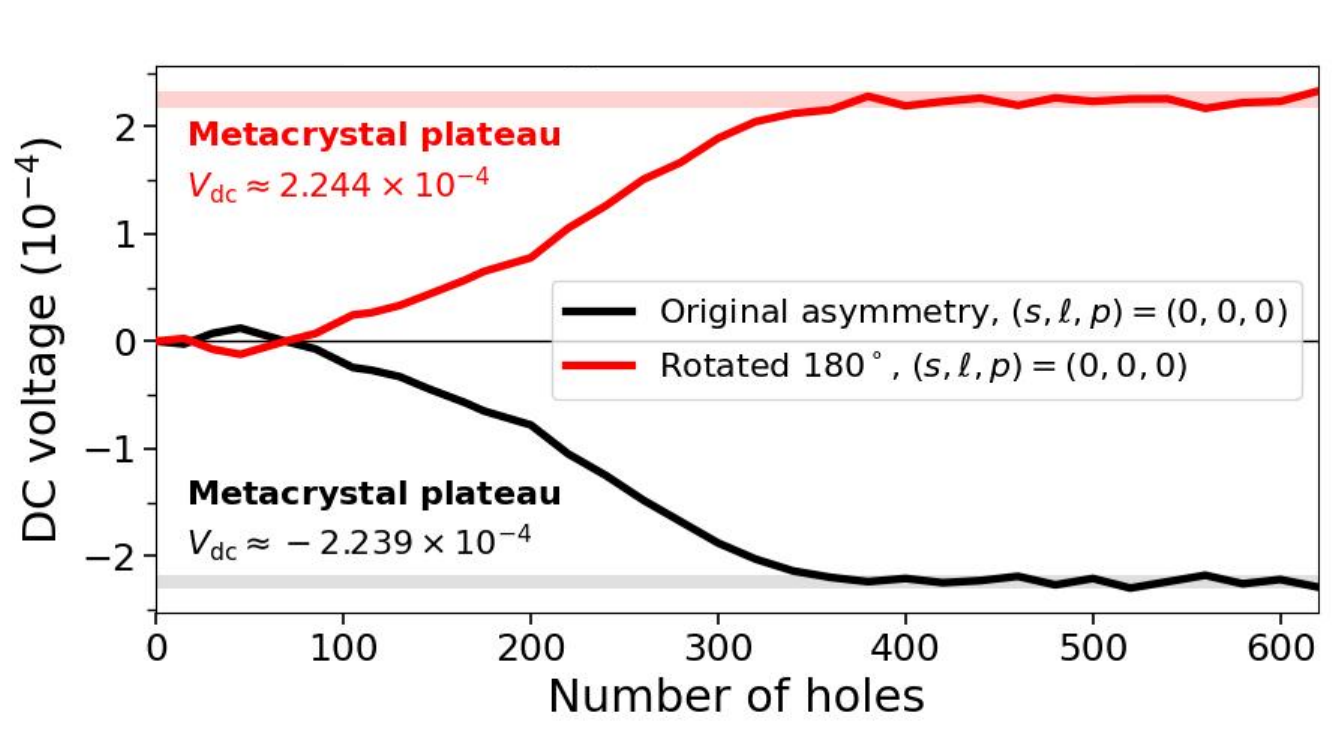}
        \put(-1.5,50){(a)}
    \end{overpic}

    \vspace{0.35em}

    \begin{overpic}[width=0.96\columnwidth]{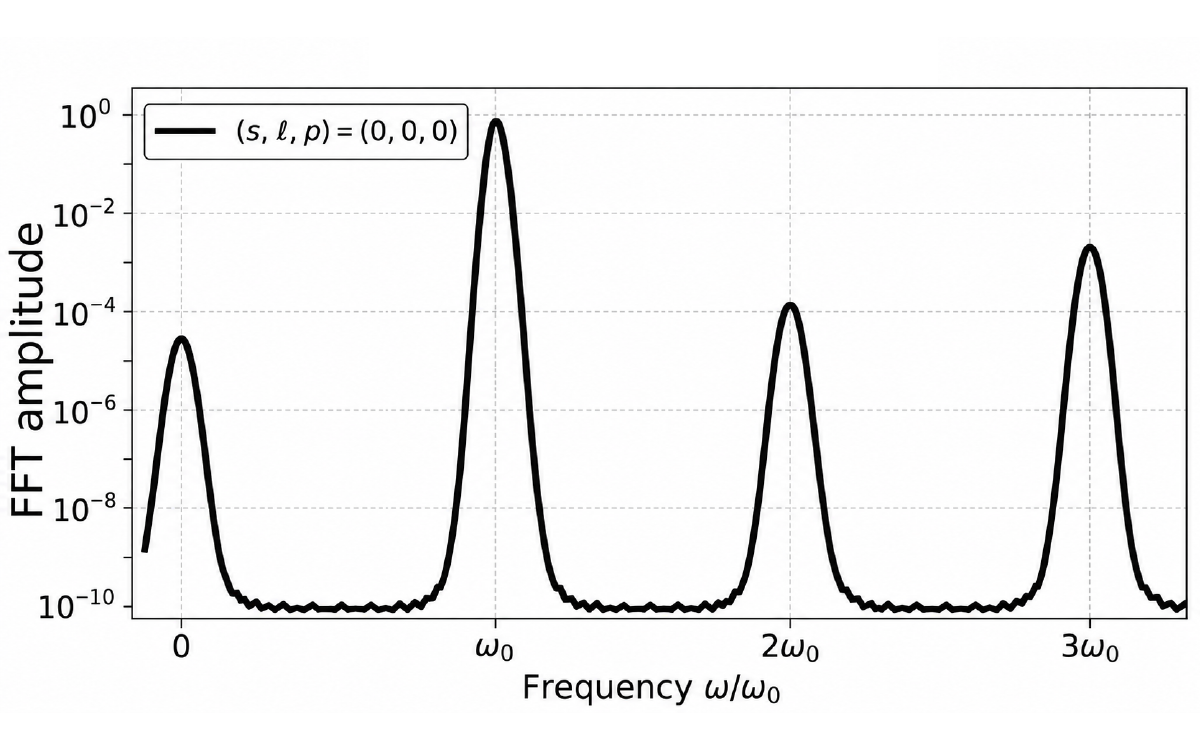}
        \put(-1.5,50){(b)}
    \end{overpic}
    
    \caption{\raggedright
Geometry dependence and harmonic content of the optically rectified voltage for
the linearly polarized Gaussian drive, \((s,\ell,p)=(0,0,0)\). Panel (a) shows
the dc voltage as the number of asymmetric holes is increased for the original
hole orientation and for the \(180^\circ\)-rotated geometry. The voltage grows
with hole number and approaches a metacrystal plateau, while rotating the holes
reverses the sign of the response, showing that the diode polarity is set by
the broken spatial symmetry of the hole array. Panel (b) shows the Fourier
spectrum of the voltage response for the 169-hole metacrystal. The spectrum
contains a finite dc component together with peaks at \(2\omega_0\), and \(3\omega_0\), indicating nonlinear voltage dynamics beyond
a purely quadratic response to the optical field.
    }
    \label{fig:dc_voltage_fft_summary}
\end{figure}

Figure~\ref{fig:dc_voltage_fft_summary}(a) shows the dc voltage as a function of the number of asymmetric holes for a Gaussian $y$-polarized drive. As the number of holes increases, the magnitude of the rectified voltage grows and then begins to saturate, showing that the response accumulates across the patterned film. Rotating the mirror axis of the holes by $180^\circ$ reverses the sign of the dc voltage under the same optical drive, demonstrating that the polarity of the rectified response is determined by the broken spatial symmetry of the hole geometry.

For 169 holes, Fig.~\ref{fig:dc_voltage_fft_summary}(b) shows the Fourier spectrum of the voltage response. The spectrum contains a finite dc component, the driven response at $\omega_0$, and higher harmonics at $2\omega_0$ and $3\omega_0$. The dc component identifies the rectified voltage, while the higher harmonics reveal nonlinear dynamics under optical driving. In particular, a purely quadratic response to the optical field would primarily generate a dc component and a second harmonic at $2\omega_0$. The prominent third harmonic at $3\omega_0$ therefore indicates that cubic and higher-order dynamical contributions are also significant.

\section{Conclusion}

We demonstrate a light-induced superconducting diode response in a patterned
superconducting film with geometrically asymmetric holes. The hole pattern
breaks left--right reflection symmetry under \(x\rightarrow -x\), allowing a
macroscopic rectified photovoltage along the \(x\) direction. Generalized TDGL
simulations produce finite photovoltages and directional supercurrent imbalance
without a conventional junction or externally applied dc bias. The line-cut
coefficients \(\eta_{\rm LR}\) and \(\eta_{\rm UD}\) characterize the internal
two-dimensional redistribution of the coherent current, and the dc
photovoltage is evaluated between the left and right probes. The response
depends on the hole array and optical mode, with circular-helicity reversal
reversing the diode polarity. Pulsed excitation enhances the zero-bias diode
coefficients beyond \(1\%\) in the 169-hole metacrystal. The present treatment
isolates the coherent condensate response and neglects optical heating and
nonequilibrium quasiparticle dynamics, which may modify the response in
experiment.

The metacrystal generates a rectified diode response for both circularly and
linearly polarized light. The linearly polarized Gaussian case deserves special
discussion because the light does not explicitly break time-reversal symmetry.
The asymmetric metacrystal mixes the driven current components, producing a
finite phase lag, nonzero current chirality, and elliptically polarized
supercurrent motion inside the patterned region. We interpret this as an
inverse-Faraday-effect-like mechanism in which structural asymmetry converts a
linearly polarized drive into local chiral current motion. Broken left--right
reflection symmetry and the nonlinear TDGL response then provide a microscopic
route to the optically induced superconducting diode effect.

Future work will focus on optimizing and extending the light-induced diode
regime using material-specific parameters. Machine-learning-guided searches
over hole shape, lattice spacing, array geometry, beam structure, pulse
profile, and material parameters could identify designs with larger zero-bias
diode coefficients and rectified photovoltages. The
inverse-Faraday-effect-like mechanism will also be analyzed more directly,
and the simulations will be extended to finite applied voltage to determine
how the zero-bias response changes with vortex nucleation, phase slips, and
magnetochiral anisotropy.

\begin{acknowledgments}

\section{Acknowledgements}
We acknowledge useful discussions with Dr. Naoto Nagaosa, Dr. Pavel Volkov,
Dr. Joshuah Heath, and Dr. Tien Tien Yeh, as well as support from DOE Quantum
Rectification SC-0025580. This work was performed in part at the Center for
Integrated Nanotechnologies, an Office of Science User Facility operated for
the U.S. Department of Energy Office of Science. Los Alamos National
Laboratory, an affirmative action equal opportunity employer, is managed by
Triad National Security, LLC for the U.S. DOE NNSA under Contract
No.~89233218CNA000001.
\end{acknowledgments}

\bibliographystyle{apsrev4-2}
\bibliography{refs}

@article{BISHOP23CPC,
  title   = {pyTDGL: Time-dependent Ginzburg-Landau in Python},
  author  = {Bishop-Van Horn, Logan},
  journal = {Computer Physics Communications},
  volume  = {291},
  pages   = {108799},
  year    = {2023},
  doi     = {10.1016/j.cpc.2023.108799}
}

@article{Matsubara2022Polarization,
  author  = {Matsubara, Masakazu and Kobayashi, Takatsugu and Watanabe, Hikaru and Yanase, Youichi and Iwata, Satoshi and Kato, Takeshi},
  title   = {Polarization-controlled tunable directional spin-driven photocurrents in a magnetic metamaterial with threefold rotational symmetry},
  journal = {Nature Communications},
  year    = {2022},
  volume  = {13},
  number  = {1},
  pages   = {6708},
  doi     = {10.1038/s41467-022-34374-7},
  url     = {https://doi.org/10.1038/s41467-022-34374-7}
}

@article{Pettine2024LightDriven,
  author  = {Pettine, Jacob and Padmanabhan, Prashant and Shi, Teng and Gingras, Lauren and McClintock, Luke and Chang, Chun-Chieh and Kwock, Kevin W. C. and Yuan, Long and Huang, Yue and Nogan, John and Baldwin, Jon K. and Adel, Peter and Holzwarth, Ronald and Azad, Abul K. and Ronning, Filip and Taylor, Antoinette J. and Prasankumar, Rohit P. and Lin, Shi-Zeng and Chen, Hou-Tong},
  title   = {Light-driven nanoscale vectorial currents},
  journal = {Nature},
  year    = {2024},
  volume  = {626},
  number  = {8001},
  pages   = {984--989},
  doi     = {10.1038/s41586-024-07037-4},
  url     = {https://doi.org/10.1038/s41586-024-07037-4}
}

@article{Yang2023LinearIFE,
  title        = {An inverse Faraday effect generated by linearly polarized light through a plasmonic nano-antenna},
  author       = {Yang, Xingyu and Mou, Ye and Zapata, Romeo and Reynier, Beno{\^{\i}}t and Gallas, Bruno and Mivelle, Mathieu},
  journal      = {Nanophotonics},
  year         = {2023},
  volume       = {12},
  number       = {4},
  pages        = {687--694},
  doi          = {10.1515/nanoph-2022-0488}
}

@article{Ando2020Nature,
  author  = {Ando, Fuyuki and Miyasaka, Yuki and Li, Tairan and Ishizuka, Jun and Arakawa, Tsutomu and Shiota, Yoichi and Moriyama, Takahiro and Yanase, Youichi and Ono, Teruo},
  title   = {Observation of superconducting diode effect},
  journal = {Nature},
  year    = {2020},
  volume  = {584},
  number  = {7821},
  pages   = {373--376},
  doi     = {10.1038/s41586-020-2590-4}
}

@article{Lyu2021NatCommun,
  author  = {Lyu, Yang-Yang and Jiang, Ji and Wang, Yong-Lei and Xiao, Zhi-Li and Dong, Sining and Chen, Qing-Hu and Milo{\v s}evi{\'c}, Milorad V. and Wang, Huabing and Divan, Ralu and Pearson, John E. and Wu, Peiheng and Peeters, Francois M. and Kwok, Wai-Kwong},
  title   = {Superconducting diode effect via conformal-mapped nanoholes},
  journal = {Nature Communications},
  year    = {2021},
  volume  = {12},
  pages   = {2703},
  doi     = {10.1038/s41467-021-23077-0}
}

@article{Nadeem2023NatRevPhys,
  author  = {Nadeem, Muhammad and Fuhrer, Michael S. and Wang, Xiaolin},
  title   = {The superconducting diode effect},
  journal = {Nature Reviews Physics},
  year    = {2023},
  volume  = {5},
  number  = {10},
  pages   = {558--577},
  doi     = {10.1038/s42254-023-00632-w}
}

@article{Daido2022PRL,
  author  = {Daido, Akito and Ikeda, Yuhei and Yanase, Youichi},
  title   = {Intrinsic Superconducting Diode Effect},
  journal = {Physical Review Letters},
  year    = {2022},
  volume  = {128},
  number  = {3},
  pages   = {037001},
  doi     = {10.1103/PhysRevLett.128.037001}
}

@article{Yuan2022PNAS,
  author  = {Yuan, Noah F. Q. and Fu, Liang},
  title   = {Supercurrent diode effect and finite-momentum superconductors},
  journal = {Proceedings of the National Academy of Sciences},
  year    = {2022},
  volume  = {119},
  number  = {15},
  pages   = {e2119548119},
  doi     = {10.1073/pnas.2119548119}
}

@article{Wakatsuki2017SciAdv,
  author  = {Wakatsuki, Ryohei and Saito, Yu and Hoshino, Shintaro and Itahashi, Yuki M. and Ideue, Toshiya and Ezawa, Motohiko and Iwasa, Yoshihiro and Nagaosa, Naoto},
  title   = {Nonreciprocal charge transport in noncentrosymmetric superconductors},
  journal = {Science Advances},
  year    = {2017},
  volume  = {3},
  number  = {4},
  pages   = {e1602390},
  doi     = {10.1126/sciadv.1602390}
}

@article{Nagaosa2024ARCMP,
  author  = {Nagaosa, Naoto and Yanase, Youichi},
  title   = {Nonreciprocal Transport and Optical Phenomena in Quantum Materials},
  journal = {Annual Review of Condensed Matter Physics},
  year    = {2024},
  volume  = {15},
  pages   = {63--83},
  doi     = {10.1146/annurev-conmatphys-032822-033734}
}

@article{Mironov2024PRB,
  author  = {Mironov, S. V. and Mel'nikov, A. S. and Buzdin, A. I.},
  title   = {Photogalvanic phenomena in superconductors supporting intrinsic diode effect},
  journal = {Physical Review B},
  year    = {2024},
  volume  = {109},
  number  = {22},
  pages   = {L220503},
  doi     = {10.1103/PhysRevB.109.L220503}
}

@article{Parafilo2025Photodiode,
  author  = {Parafilo, A. V. and Sun, Meng and Sonowal, K. and Kovalev, V. M. and Savenko, I. G.},
  title   = {Proposal for superconducting photodiode},
  journal = {2D Materials},
  volume  = {12},
  number  = {1},
  pages   = {011001},
  year    = {2025},
  doi     = {10.1088/2053-1583/ad9596}
}

@article{Yeh2025QuantumPrintingI,
  author  = {Yeh, Tien-Tien and Yerzhakov, Hennadii and Bishop-Van Horn, Logan and Raghu, Srinivas and Balatsky, A. V.},
  title   = {Quantum printing and induced vorticity in superconductors I: Linearly polarized light},
  journal = {Physical Review Research},
  volume  = {7},
  number  = {4},
  pages   = {043111},
  year    = {2025},
  doi     = {10.1103/k9m4-h474}
}

@article{Yeh2025QuantumPrintingII,
  author  = {Yeh, Tien-Tien and Yerzhakov, Hennadii and Bishop-Van Horn, Logan and Raghu, Srinivas and Balatsky, A. V.},
  title   = {Quantum printing and induced vorticity in superconductors II: Laguerre-Gaussian beam},
  journal = {Physical Review Research},
  volume  = {7},
  number  = {4},
  pages   = {043112},
  year    = {2025},
  doi     = {10.1103/dqv7-w2w4}
}

@article{Aeppli2025QuantumPrinting,
  author  = {Aeppli, Gabriel and Balatsky, Alexander V. and Bonetti, Stefano
             and Cardoso, Gabriel and Raghu, Srinivas and Sylju{\aa}sen, Erlend
             and Yeh, Tien-Tien and Lin, Shi-Zeng and Liu, Yuefei
             and Weissenrieder, Jonas and Wong, Patrick J.},
  title   = {Quantum Printing},
  journal = {\href{https://arxiv.org/abs/2509.16792}{arXiv:2509.16792 [quant-ph]}},
  year    = {2025}
}

@article{Baumgartner2022JosephsonDiode,
  author  = {Baumgartner, C. and Fuchs, L. and Costa, A. and Reinhardt, S. and Gronin, S. and Gardner, G. C. and Lindemann, T. and Manfra, M. J. and Faria Junior, P. E. and Kochan, D. and Fabian, J. and Paradiso, N. and Strunk, C.},
  title   = {Supercurrent rectification and magnetochiral effects in symmetric Josephson junctions},
  journal = {Nature Nanotechnology},
  volume  = {17},
  pages   = {39--44},
  year    = {2022},
  doi     = {10.1038/s41565-021-01009-9}
}

@article{KramerWattsTobin1978,
  author  = {Kramer, Lorenz and Watts-Tobin, R. J.},
  title   = {Theory of dissipative current-carrying states in superconducting filaments},
  journal = {Physical Review Letters},
  volume  = {40},
  pages   = {1041--1044},
  year    = {1978},
  doi     = {10.1103/PhysRevLett.40.1041}
}

@article{VanDerZiel1965,
  author  = {van der Ziel, J. P. and Pershan, P. S. and Malmstrom, L. D.},
  title   = {Optically-Induced Magnetization Resulting from the Inverse Faraday Effect},
  journal = {Physical Review Letters},
  volume  = {15},
  pages   = {190--193},
  year    = {1965},
  doi     = {10.1103/PhysRevLett.15.190}
}

@article{Mironov2021IFECondensates,
  author  = {Mironov, S. V. and Mel'nikov, A. S. and Tokman, I. D. and Vadimov, V. and Lounis, B. and Buzdin, A. I.},
  title   = {Inverse Faraday Effect for Superconducting Condensates},
  journal = {Physical Review Letters},
  volume  = {126},
  number  = {13},
  pages   = {137002},
  year    = {2021},
  doi     = {10.1103/PhysRevLett.126.137002}
}

@article{Ciaccia2023GateTunableJD,
  title = {Gate-tunable Josephson diode in proximitized {InAs} supercurrent interferometers},
  author = {Ciaccia, Carlo and Haller, Roy and Drachmann, Asbj{\o}rn C. C. and Lindemann, Tyler and Manfra, Michael J. and Schrade, Constantin and Sch{\"o}nenberger, Christian},
  journal = {Physical Review Research},
  volume = {5},
  pages = {033131},
  year = {2023},
  doi = {10.1103/PhysRevResearch.5.033131}
}

@article{Zeng2016AlOxBarrier,
  title = {Atomic structure and oxygen deficiency of the ultrathin aluminium oxide barrier in {Al/AlOx/Al} Josephson junctions},
  author = {Zeng, Lunjie and Tran, Dung Trung and Tai, Cheuk-Wai and Svensson, Gunnar and Olsson, Eva},
  journal = {Scientific Reports},
  volume = {6},
  pages = {29679},
  year = {2016},
  doi = {10.1038/srep29679}
}

@article{Wang2025PerfectQSD,
  title = {Quantum superconducting diode effect with perfect efficiency above liquid-nitrogen temperature},
  author = {Wang, Heng and Zhu, Yuying and Bai, Zhonghua and Lyu, Zhaozheng and Yang, Jiangang and Zhao, Lin and Zhou, X. J. and Gu, Genda and Xue, Qi-Kun and Zhang, Ding},
  journal = {Nature Physics},
  volume = {22},
  pages = {47--53},
  year = {2026},
  doi = {10.1038/s41567-025-03098-y}
}

@article{Cordoso2026PRL,
  title = {Orbital Inverse Faraday and Cotton-Mouton Effects in Hall Fluids},
  author = {Cardoso, Gabriel and Sylju\aa{}sen, Erlend and Balatsky, Alexander V.},
  journal = {Phys. Rev. Lett.},
  volume = {136},
  issue = {1},
  pages = {016502},
  numpages = {6},
  year = {2026},
  month = {Jan},
  publisher = {American Physical Society},
  doi = {10.1103/z4db-gt69},
  url = {https://link.aps.org/doi/10.1103/z4db-gt69}
}

@article{Suri2022APL,
  author  = {Suri, Dhavala and Kamra, Akashdeep and Meier, Thomas N. G. and Kronseder, Matthias and Belzig, Wolfgang and Back, Christian H. and Strunk, Christoph},
  title   = {Non-reciprocity of vortex-limited critical current in conventional superconducting micro-bridges},
  journal = {Applied Physics Letters},
  volume  = {121},
  number  = {10},
  pages   = {102601},
  year    = {2022},
  doi     = {10.1063/5.0109753}
}

@article{Pal2022NatPhys,
  author  = {Pal, Banabir and Chakraborty, Anirban and Sivakumar, Pranava K. and Davydova, Margarita and Gopi, Ajesh K. and Pandeya, Avanindra K. and Krieger, Jonas A. and Zhang, Yang and Date, Mihir and Ju, Sailong and Yuan, Noah and Schr{\"o}ter, Niels B. M. and Fu, Liang and Parkin, Stuart S. P.},
  title   = {Josephson diode effect from {Cooper} pair momentum in a topological semimetal},
  journal = {Nature Physics},
  volume  = {18},
  number  = {10},
  pages   = {1228--1233},
  year    = {2022},
  doi     = {10.1038/s41567-022-01699-5}
}

@article{Chen2018PRB,
  author  = {Chen, Chui-Zhen and He, James Jun and Ali, Mazhar N. and Lee, Gil-Ho and Fong, Kin Chung and Law, K. T.},
  title   = {Asymmetric {Josephson} effect in inversion symmetry breaking topological materials},
  journal = {Physical Review B},
  volume  = {98},
  number  = {7},
  pages   = {075430},
  year    = {2018},
  doi     = {10.1103/PhysRevB.98.075430}
}

@article{Legg2022PRB,
  author  = {Legg, Henry F. and Loss, Daniel and Klinovaja, Jelena},
  title   = {Superconducting diode effect due to magnetochiral anisotropy in topological insulators and {Rashba} nanowires},
  journal = {Physical Review B},
  volume  = {106},
  number  = {10},
  pages   = {104501},
  year    = {2022},
  doi     = {10.1103/PhysRevB.106.104501}
}

@article{jonsson2022current,
  title   = {Current Crowding in Nanoscale Superconductors within the Ginzburg--Landau Model},
  author  = {J{\"o}nsson, Mattias and Vedin, Robert and Gyger, Samuel and Sutton, James A. and Steinhauer, Stephan and Zwiller, Val and Wallin, Mats and Lidmar, Jack},
  journal = {Physical Review Applied},
  volume  = {17},
  number  = {6},
  pages   = {064046},
  year    = {2022},
  month   = {jun},
  doi     = {10.1103/PhysRevApplied.17.064046}
}

@article{Edelstein1995PRL,
  author  = {Edelstein, Victor M.},
  title   = {Magnetoelectric Effect in Polar Superconductors},
  journal = {Physical Review Letters},
  volume  = {75},
  number  = {10},
  pages   = {2004--2007},
  year    = {1995},
  month   = {sep},
  doi     = {10.1103/PhysRevLett.75.2004}
}

@article{Edelstein1996JPCM,
  author  = {Edelstein, Victor M.},
  title   = {The Ginzburg--Landau Equation for Superconductors of Polar Symmetry},
  journal = {Journal of Physics: Condensed Matter},
  volume  = {8},
  number  = {3},
  pages   = {339--349},
  year    = {1996},
  month   = {jan},
  doi     = {10.1088/0953-8984/8/3/012}
}

@article{Wei2020ZeroBiasMIR,
  author  = {Wei, Jingxuan and Li, Ying and Wang, Lin and Liao, Wugang and Dong, Bowei and Xu, Cheng and Zhu, Chunxiang and Ang, Kah-Wee and Qiu, Cheng-Wei and Lee, Chengkuo},
  title   = {Zero-bias mid-infrared graphene photodetectors with bulk photoresponse and calibration-free polarization detection},
  journal = {Nature Communications},
  volume  = {11},
  pages   = {6404},
  year    = {2020},
  doi     = {10.1038/s41467-020-20115-1}
}

@article{Wei2023GeometricMIR,
  author  = {Wei, Jingxuan and Chen, Yang and Li, Ying and Li, Wei and Xie, Junsheng and Lee, Chengkuo and Novoselov, Kostya S. and Qiu, Cheng-Wei},
  title   = {Geometric filterless photodetectors for mid-infrared spin light},
  journal = {Nature Photonics},
  volume  = {17},
  pages   = {171--178},
  year    = {2023},
  doi     = {10.1038/s41566-022-01115-7}
}

@misc{MetaSDEVideos2026,
  author       = {{Balatsky Group}},
  title        = {Metacrystal superconducting diode effect simulation videos},
  year         = {2026},
  howpublished = {\url{https://sites.google.com/balatskygroup.org/web/metasde}},
  note         = {Accessed August 11, 2026}
}

\appendix

\setcounter{secnumdepth}{1}

\section{Numerical implementation and observables}
\label{app:numerics}

\begin{figure*}[t]
    \centering
    \includegraphics[width=0.98\textwidth]{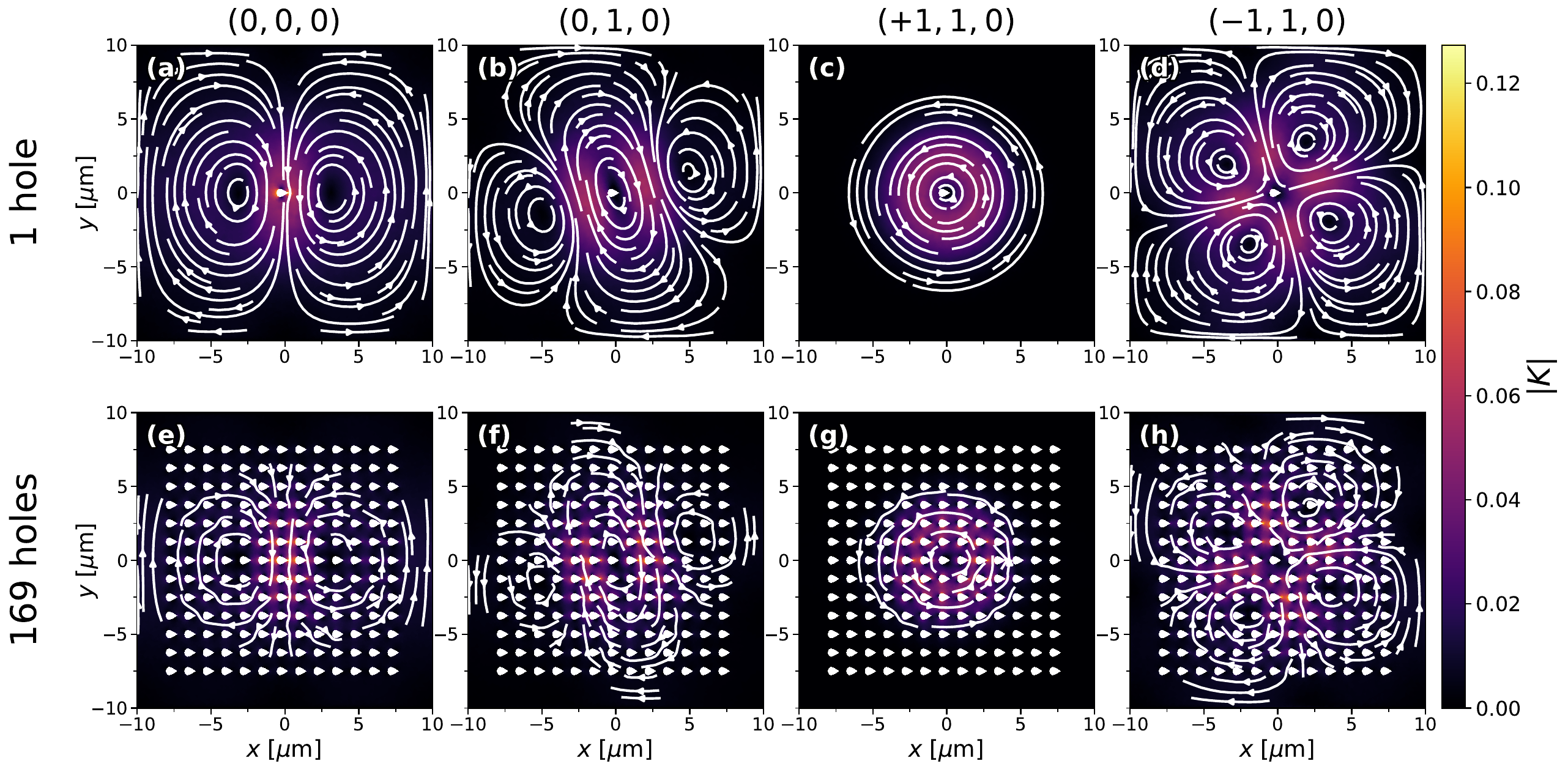}
    \caption{\raggedright
Optically induced sheet-current textures for different optical modes and
asymmetric-hole arrays. Panels (a)--(d) show one hole and panels (e)--(h)
the 169-hole metacrystal. Columns compare the linearly polarized Gaussian
\((s,\ell,p)=(0,0,0)\), linearly polarized Laguerre--Gaussian \((0,1,0)\),
and opposite-helicity circular Laguerre--Gaussian modes \((+1,1,0)\) and
\((-1,1,0)\). The color scale gives \(|\mathbf K|\) and streamlines the local
current direction. Increasing hole number redistributes current across the
metacrystal, while optical mode and helicity reshape the current texture and
the orientation of the chiral flow. Video simulations for all optical modes are available on the companion website~\cite{MetaSDEVideos2026}.
    }
    \label{fig:current_textures}
\end{figure*}

In dimensionless units with \(\sigma=1\),
\(\mathbf J_n=-\nabla\mu-\partial_t\mathbf A\), and the electrochemical
potential is determined self-consistently from charge conservation,
\begin{equation}
\nabla\cdot\mathbf J=0,
\qquad
\nabla^2\mu=
\nabla\cdot\mathbf J_s-\nabla\cdot\partial_t\mathbf A .
\end{equation}

All outer film edges and hole boundaries are insulating
superconductor--vacuum interfaces~\cite{jonsson2022current}, with
\begin{equation}
\hat{\mathbf n}\cdot(\nabla-i\mathbf A)\psi=0,
\end{equation}
so that \(\hat{\mathbf n}\cdot\mathbf J_s=0\). The corresponding condition
\(\hat{\mathbf n}\cdot(\nabla\mu+\partial_t\mathbf A)=0\) gives
\(\hat{\mathbf n}\cdot\mathbf J_n=0\).

We use \(d=0.02~\mu{\rm m}\), \(\xi=\lambda=0.1~\mu{\rm m}\),
\(\rho_n=150~\mu\Omega\,{\rm cm}\), and a square hole lattice with
\(a_x=a_y=1.25~\mu{\rm m}\). For Fig.~\ref{fig:dc_voltage_fft_summary}(a),
the \(40~\mu\mathrm{m}\times40~\mu\mathrm{m}\) film contains 0--600 holes
at fixed geometry and spacing. The characteristic mesh spacing is \(\xi/4\), resolving the order-parameter amplitude, phase gradients, and current density near the hole boundaries; the diode coefficients are unchanged within numerical precision for \(\xi/3\) and finer meshes, increasing the simulation domain produces no qualitative change, and the current-imbalance coefficients vanish without optical driving.

The TDGL time scale is
\(\tau_0=\mu_0\sigma\lambda^2=8.38~\mathrm{fs}\), giving
\(f=11.9~\mathrm{THz}\); \(E_0=0.5\) corresponds to a peak Gaussian
electric field of approximately \(0.60~\mathrm{kV\,cm^{-1}}\). Pulsed
simulations use
\(g(t)=\exp[-(t-t_0)^2/(2\sigma_t^2)]\), with \(t_0=50\),
\(\sigma_t=10\), \(E_0=0.5\), and \(w_0=3~\mu\mathrm{m}\).

The mode labels are \((0,0,0)\) for linearly \(y\)-polarized Gaussian,
\((0,1,0)\) for linearly \(y\)-polarized Laguerre--Gaussian with
\((\ell,p)=(1,0)\), \((\pm1,0,0)\) for circular Gaussian, and
\((\pm1,1,0)\) for circular Laguerre--Gaussian. Circular \(x\) and \(y\)
components differ in phase by \(\pi/2\), and the sign of \(s\) sets the
helicity. The same \(E_0\) is used for all polarizations without
\(1/\sqrt{2}\) rescaling of the circular Cartesian components, so fixed
\(E_0\) does not imply equal cycle-averaged \(|\mathbf E|^2\).

An \(x\)-polarized Gaussian drive shows the same increase in directional
current imbalance with hole number as the \(y\)-polarized case, but with
consistently smaller magnitude; its dc voltage has a similar nonlinear
intensity dependence.

The voltage convention and dc average are
\begin{equation}
\begin{aligned}
V_{ij}(t)&=\mu(\mathbf r_i,t)-\mu(\mathbf r_j,t),\\
V_{ij}^{\mathrm{dc}}
&=\frac{1}{t_b-t_a}\int_{t_a}^{t_b}V_{ij}(t)\,dt .
\end{aligned}
\end{equation}
The left and right probes are at
\((x,y)=(-9~\mu{\rm m},0)\) and \((9~\mu{\rm m},0)\) in the unpatterned
regions outside the metacrystal, with probe ordering fixing the voltage sign.
For continuous driving, the post-transient average spans an integer number
of optical periods, so
\(\mathbf A(\mathbf r,t_b)=\mathbf A(\mathbf r,t_a)\) and the cycle average
of \(\partial_t\int_{\mathcal C_{j\rightarrow i}}\mathbf A\cdot d\mathbf l\)
vanishes. Hence the reported dc voltage equals
\(\langle\mu(\mathbf r_i,t)-\mu(\mathbf r_j,t)\rangle\).

\section{Extended optical-mode dependence}
\label{app:modes}

For the Gaussian mode, the similar \(\eta_{\rm LR}\) and
\(\eta_{\rm UD}\) values in Fig.~\ref{fig:eta_phase_lag}(a) reflect
comparable redistribution through the two central cuts. Their asymmetric
positive and negative steady-state peaks do not cancel over a cycle, yielding
a finite zero-bias average. Table~\ref{tab:eta_modes} gives
\(\eta_{\rm UD}=0.54\%\) for the 169-hole linearly polarized
Laguerre--Gaussian drive, compared with \(-0.24\%\) for the Gaussian drive;
opposite circular helicities reverse the sign for both Gaussian and
Laguerre--Gaussian modes. Pulsing raises several 169-hole coefficients from
sub-percent values to about or above \(1\%\) while preserving the
mode-dependent structure: the pulsed Gaussian coefficients are negative in
both channels, Gaussian-helicity reversal flips both signs, the linearly
polarized Laguerre--Gaussian pulse gives the largest up--down response,
\(\eta_{\rm UD}=1.28\%\), and circular Laguerre--Gaussian pulses retain
helicity-dependent sign reversal. The enhancement reflects stronger transient
order-parameter and current-redistribution dynamics.

Figure~\ref{fig:current_textures} shows that the single-hole current extends
across the film, while the 169-hole lattice distributes the geometric
asymmetry throughout the illuminated region and repeatedly redirects the
flow. Orbital angular momentum redistributes this flow for the linearly
polarized Laguerre--Gaussian mode, while circular Laguerre--Gaussian modes
produce helicity-dependent chiral textures. The streamlines therefore show
how optical mode controls where current is enhanced and how it circulates
around the asymmetric holes.

\end{document}